\documentclass[twocolumn,showpacs,pre,superscriptaddress]{revtex4}

\usepackage{graphicx}
\usepackage{color}
\usepackage{tabularx}
\usepackage{epsfig}
\usepackage{amsmath}
\usepackage{amssymb}
\usepackage{graphicx}
\usepackage{dcolumn}
\usepackage{bm}
\usepackage{wasysym}
\usepackage{times}

\definecolor{grn}{rgb}{0,0,0.54}

\newcommand{\bra}[1]{\langle #1|}
\newcommand{\ket}[1]{|#1\rangle}
\newcommand{\braket}[2]{\langle #1|#2\rangle}

\newcommand{\qdmlr}{\unitlength 0.03em
  \begin{minipage}{35\unitlength}
    \begin{center}
      \begin{picture}(30,17)
        \put(0,0){\linethickness{0.20mm}\line(1,0){20}}
	\put(10,17.3){\linethickness{0.20mm}\line(1,0){20}}
      \end{picture}
    \end{center}
  \end{minipage}
}

\newcommand{\qdmud}{\unitlength 0.03em 
  \begin{minipage}{35\unitlength}
    \begin{center}
      \begin{picture}(30,17)
        \linethickness{0.20mm}\qbezier(0,0)(5,8.65)(10,17.3)
        \linethickness{0.20mm}\qbezier(20,0)(25,8.65)(30,17.3)
      \end{picture}
    \end{center}
  \end{minipage}
}

\newcommand{\plaqueteA}{\unitlength 0.02em 
  \begin{minipage}{35\unitlength}
    \begin{center}
      \begin{picture}(30,17)
        \linethickness{0.14mm}\qbezier(0,0)(5,8.65)(10,17.3)
        \linethickness{0.14mm}\qbezier(20,0)(25,8.65)(30,17.3)
        \put(0,0){\linethickness{0.14mm}\line(1,0){20}}
        \put(10,17.3){\linethickness{0.14mm}\line(1,0){20}}
      \end{picture}
    \end{center}
  \end{minipage}
}

\newcommand{\plaqueteB}{\unitlength 0.02em 
  \begin{minipage}{35\unitlength}
    \begin{center}
      \begin{picture}(30,17)
        \linethickness{0.14mm}\qbezier(10,0)(5,8.65)(0,17.3)
        \linethickness{0.14mm}\qbezier(30,0)(25,8.65)(20,17.3)
        \put(10,0){\linethickness{0.14mm}\line(1,0){20}}
        \put(0,17.3){\linethickness{0.14mm}\line(1,0){20}}
      \end{picture}
    \end{center}
  \end{minipage}
}

\newcommand{\plaqueteC}{\unitlength 0.02em 
  \begin{minipage}{35\unitlength}
    \begin{center}
      \begin{picture}(30,35)
        \linethickness{0.14mm}\qbezier(10,0)(5,8.65)(0,17.3)
        \linethickness{0.14mm}\qbezier(20,17.3)(20,17.3)(10,34.6)
        \linethickness{0.14mm}\qbezier(0,17.3)(0,17.3)(10,34.6)
        \linethickness{0.14mm}\qbezier(10,0)(15,8.65)(20,17.3)
      \end{picture}
    \end{center}
  \end{minipage}
}

\begin{document}

\title{ENCORE: An Extended Contractor Renormalization algorithm}

\author{A.~Fabricio Albuquerque}
\affiliation{Theoretische Physik, ETH Zurich, 8093 Zurich, Switzerland}
\affiliation{School of Physics, The University of New South Wales, Sydney, NSW 2052, Australia}

\author{Helmut G.~Katzgraber}
\affiliation{Theoretische Physik, ETH Zurich, 8093 Zurich, Switzerland}
\affiliation{Department of Physics, Texas A\&M University, College Station,
             Texas 77843-4242, USA}

\author{Matthias Troyer}
\affiliation{Theoretische Physik, ETH Zurich, 8093 Zurich, Switzerland}

\date{\today}

\begin{abstract}

Contractor renormalization (CORE) is a real-space renormalization-group
method to derive effective Hamiltionians for microscopic models. The
original CORE method is based on a real-space decomposition of the
lattice into small blocks and the effective degrees of freedom on the
lattice are tensor products of those on the small blocks. We present
an extension of the CORE method that overcomes this restriction. Our
generalization allows the application of CORE to derive arbitrary
effective models whose Hilbert space is not just a tensor product
of local degrees of freedom. The method is especially well suited
to search for microscopic models to emulate low-energy exotic models
and can guide the design of quantum devices.

\end{abstract}

\pacs{87.55.kd,03.67.Ac,02.70.-c,03.67.Pp}
\maketitle

\section{Introduction}
\label{sec:introduction}

Identifying the emergent low-energy degrees of freedom in
a strongly-correlated system is a highly nontrivial problem
requiring considerable physical intuition and a careful analysis
of available experimental data \cite{laughlin:00}. The contractor
renormalization (CORE) method introduced by Morningstar and Weinstein
\cite{morningstar:94,morningstar:96} is a tool to systematically
perform this task: by suitably selecting low-energy local degrees of
freedom and applying a real-space renormalization procedure, one can
in principle obtain an effective Hamiltonian which is simpler than
the original one and therefore (ideally) more amenable to subsequent
analytical or numerical treatment. For recent applications of CORE
see, for example, Refs.~\onlinecite{siu:07,abendschein:07,siu:08,
albuquerque:08c,picon:08,abendschein:08}.

The idea behind CORE is to divide the lattice on which the
model is defined into blocks and to retain only a small number of
suitably chosen low-lying block eigenstates. The low-energy eigenstates
of the full Hamiltonian defined on a cluster formed by two or more
blocks are then projected onto the restricted basis formed by tensor
products of the retained block states. By requiring that the low-energy
spectrum of the full problem is exactly reproduced, an effective
Hamiltonian is obtained. The mapping onto a coarser-grained lattice
with redefined degrees of freedom is done at the expense of having
longer-range effective interactions.  A successful application of the
method relies on a fast decay of the effective interactions, which in
turn depends on the correct choice of the effective degrees of freedom
and on the particular way the lattice is divided into blocks, as well
as on how the retained block eigenstates are chosen.  Because physical
intuition and a good idea of the relevant local degrees of freedom
are needed to obtain physically sound results, we believe that this
is the main reason why CORE has not found a more widespread use.

The ``inverse'' problem of using CORE to find microscopic models
that map well onto a desired effective low-energy Hamiltonian does
not suffer from the aforementioned problems: because the emergent
degrees of freedom are known {\em a priori} and their adequacy in
describing the low-energy physics of the device is {\em enforced}
by design, the aforementioned limitations of CORE can be used to
our advantage. Whenever the effective Hamiltonian includes sizable
long-range interactions and/or states with a vanishing projection
on the restricted basis are present in the low-energy spectrum,
one can conclude that the considered microscopic model is not well
approximated by the proposed low-energy effective model.  Given current
interest in the emulation of exotic phases via physical models
(e.g., by using Josephson junctions or cold atomic/molecular gases),
we expect this approach to be useful when designing manipulable {\em
quantum tool boxes}.  Finally, we note that the step of dividing the
lattice into blocks is no longer required or even desirable within
this context and we thus extend the method to models built from
geometrically-constrained degrees of freedom, such as quantum dimer
models \cite{rokhsar:88, moessner:01}, where the emergent degrees
of freedom cannot be described in terms of tensor products of local
states. Below we introduce an extension of the CORE method applicable
to arbitrary basis states of the effective model and illustrate the
application of the method with an array of quantum Josephson junctions
\cite{ioffe:02} used to implement a topologically protected qubit
\cite{albuquerque:08}.

\section{Extended CORE method}
\label{sec:core}

We first review the standard CORE algorithm
\cite{morningstar:94,morningstar:96,altman:02,capponi:06} and then
contrast it to the extended CORE (dubbed ENCORE) method proposed here.

\subsection{Standard CORE algorithm}
\label{sec:core.standard}

Given a Hamiltonian ${\mathcal H}$ defined on a lattice ${\mathcal L}$,
the standard CORE algorithm can be described as follows:

\begin{enumerate}

\item{Divide the lattice ${\mathcal L}$ into disconnected small blocks
${\mathcal B}$ and diagonalize the Hamiltonian ${\mathcal H}$ within
a single block, keeping $M$ low-lying eigenstates $\{ \ket{\phi_m}
\}_{1}^{M}$. The subspace spanned by tensor products of these block
eigenstates on a cluster ${\mathcal C}$---formed by joining a number
of elementary blocks---defines the reduced Hilbert space within which
the effective model is derived.}

\item{Diagonalize ${\mathcal H}$ on a cluster ${\mathcal C}$
consisting of $N$ connected blocks retaining the ${\mathcal M} =
M^{N}$ lowest eigenstates $\{\ket{n} \}_{1}^{\cal{M}}$ with energies
$\epsilon_n$ and project them onto the basis formed by the tensor
products of the block states, $\{ \ket{\phi_{m_1}, ... , \phi_{m_{N}}}
\}_{1}^{\cal{M}}$, forming the projected states $\{ \ket{\psi_n}
\}_{1}^{\cal{M}}$.}

\item{ Orthonormalize the obtained projected states $\{ \ket{\psi_n}
\}_{1}^{\cal{M}}$ using a Gramm-Schmidt procedure
\begin{equation}
\ket{\tilde{\psi}_n} = {\frac{1}{Z_n}} \bigg( \ket{\psi_n} - 
\sum_{m<n} \ket{\tilde{\psi}_m} \braket{\tilde{\psi}_m}  {\psi_n}\bigg) ,
\label{eq:gramm}
\end{equation}
where $Z_n$ stands for the normalization of the orthogonalized state.}

\item{The range-$N$ renormalized Hamiltonian is then 
\begin{equation}
{\mathcal H}^{\rm ren}_N =  \sum_{n}^{\cal{M}} \epsilon_n \ket{\tilde{\psi}_n} 
\bra{\tilde{\psi}_n} .
\label{eq:hren}
\end{equation}
By construction, this Hamiltonian has the same low-energy spectrum
as the original one.}

\item{Writing Eq.~({\ref{eq:hren}}) in terms of the tensor product
states $\{ \ket{\phi_{m_1}, ... , \phi_{m_N}} \}_{1}^{\cal{M}}$, we
obtain the range-$N$ effective interactions between the blocks
forming the cluster after subtracting the previously calculated
shorter-range interactions 
\begin{equation}
h_{i_{1} ... i_{N}} =  
{\mathcal H}^{\rm ren}_N - \sum_{N' = 1}^{N-1} \sum_{\langle i_{1}, ... ,i_{N'} \rangle} 
h_{i_{1} ... i_{N'}} ,
\label{eq:rangeN}
\end{equation}
where $\langle i_{1}, ... ,i_{N'} \rangle$ denotes the set of {\em all
connected} range-$N'$ subclusters. The effective range-$N$
Hamiltonian can then be written as
\begin{equation}
{\mathcal H}^{\rm eff}_N =  \sum_{i}h_i + \sum_{\langle i,j \rangle} h_{ij} + 
\sum_{\langle i,j,k \rangle} h_{ijk} + \ldots ,
\label{eq:heff}
\end{equation}
where $h_i$ is the block self-energy, $h_{ij}$ the interaction
between nearest-neighbor blocks, $h_{ijk}$ a three-block coupling,
etc.~up to range-$N$ interactions.}

\end{enumerate}
The successful application of the above procedure relies on a
fast decay of the long-ranged effective interactions appearing in
Eq.~(\ref{eq:heff}) and therefore one chooses the restricted set of
degrees of freedom by specifying the elementary blocks ${\mathcal B}$
and the retained block states $\{ \ket{\phi_m}\}_{1}^{M}$.

\subsection{ENCORE algorithm}
\label{sec:core.generalized}

It is possible to extend the ideas presented in Sec.~\ref{sec:core} to
constrained effective models---e.g., quantum dimer models, loop models,
and string nets---for which the relevant degrees of freedom are no
longer formed by tensor products of block states but, instead, by the
set of configurations on a given cluster satisfying the constraints of
the Hamiltonian to be emulated. We thus present an extended algorithm
using alternative ways of selecting the restricted degrees of freedom.

\begin{enumerate}

\item{Choose a finite-size cluster ${\mathcal C}$ and build a basis
$\{ \ket{\phi_m} \}_{1}^{M}$ for the Hilbert space of the effective
model. In the standard CORE method this effective basis is a tensor
product of the relevant states on the blocks, whereas here it is
comprised by all constrained configurations on ${\mathcal C}$. For
example, for a quantum dimer model we generate all $M$ possible dimer
coverings on the cluster ${\mathcal C}$.}

\item{Diagonalize ${\mathcal H}$ on the cluster  ${\mathcal C}$,
calculating the $M$ lowest eigenstates $\{ \ket{n} \}_{1}^{M}$
with energies $\epsilon_n$ and project them onto the restricted
basis $\{\ket{\phi_m} \}_{1}^{M}$, forming the projected states
$\{\ket{\psi_m} \}_{1}^{M}$ \cite{comment:solver}.}

\item{Orthonormalize by means of a Gramm-Schmidt procedure as in
Eq.~(\ref{eq:gramm}).}

\item{The Hamiltonian within the restricted space is then given by
Eq. (\ref{eq:hren}).}

\item{Writing this Hamiltonian in the restricted basis $\{ \ket{\phi_m}
\}_{1}^{M}$ we obtain the effective model
\begin{equation}
{\mathcal H}^{\rm eff} =  \sum_{m, m', n}^{M} \epsilon_n \ket{\phi_m} 
\braket{\phi_m}{\tilde{\psi}_n} 
\braket{\tilde{\psi}_n}{\phi_{m'}}\bra{\phi_{m'}} . 
\label{eq:eff2}
\end{equation}
It is again possible to perform a cluster expansion within ENCORE by
using Eqs.~(\ref{eq:rangeN}) and (\ref{eq:heff}). }

\end{enumerate}
Note that the above discussion is for an orthonormal restricted
basis $\{\ket{\phi_m} \}_{1}^{M}$, such as in the example discussed
in Sec.~\ref{sec:application}.  Small changes in the procedure are
required if this is not the case \cite{rokhsar:88}. 

\section{Application: Emulation of the Quantum Dimer Model}
\label{sec:application}

\subsection{Array of quantum Josephson junctions}
\label{sec:JJK}

We apply the algorithm described in Sec.~\ref{sec:core.generalized}
to extract the dimer flip amplitude $t$ for a Josephson-junction array
introduced by Ioffe {\em et al.} \cite{ioffe:02} to emulate a quantum
dimer model (QDM) \cite{rokhsar:88} on a triangular lattice. This
model---first investigated by Moessner and Sondhi \cite{moessner:01}---has
the desired properties needed to implement a topologically protected
qubit and is given by ${\mathcal H} = {\mathcal H}_{\plaqueteA}
+ {\mathcal H}_{\plaqueteB} + {\mathcal H}_{\plaqueteC}$ with
\begin{equation}
\begin{split}
{{\mathcal H}_{\plaqueteA}} = 
    -t \sum_{\plaqueteA} \big[
	\ket{\qdmlr} \bra{\qdmud} + 
	\ket{\qdmud} \bra{\qdmlr}
    \big] \\
    +v \sum_{\plaqueteA} \big[
	\ket{\qdmlr} \bra{\qdmlr}  + 
	\ket{\qdmud} \bra{\qdmud}
    \big] , 
\end{split}
\label{eq:qdm_plaq}
\end{equation}
with similar definitions for ${\mathcal H}_{\plaqueteB}$ and ${\mathcal
H}_{\plaqueteC}$.  Parallel dimers on the same rhombus (henceforth
we refer to such configurations as {\em flippable rhombi}) flip
with an amplitude $t$ and interact via a potential strength $v$;
the sum runs over all rhombi with a given orientation.  Moessner and
Sondhi showed that a topologically ordered liquid phase exists
over a finite region of the model's phase diagram ($0.82 \lesssim
v/t \le 1$), something confirmed in a number of subsequent studies
\cite{ioffe:02,ralko:05,vernay:06,ralko:06}.

\begin{figure}[!tbp]
\includegraphics*[width=0.31\textwidth,angle=270]{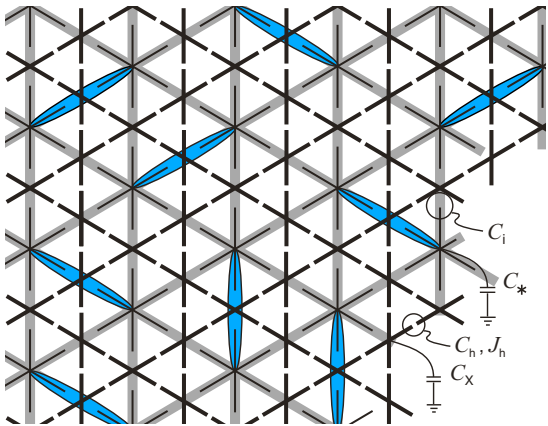}
\caption{
(Color online)
Josephson junction array used to emulate the quantum dimer model on
a triangular lattice (shaded lines, ellipses represent the dimers;
see Refs.~\cite{ioffe:02} and \cite{albuquerque:08} for details).
X-shaped superconducting islands (thick black lines) form a kagome
lattice with normal-state star-shaped islands (thin black lines) placed
at the center of every hexagon of the kagome lattice.  Cooper pairs
hop between nearest-neighbor X-shaped islands with an amplitude
given by the Josephson current $J_{\rm h}$. A large ratio between
the capacitances $C_{\rm i}$ and $C_{\rm h}$ ensures an on-hexagon
repulsion $E_{\rm hex}$ to emulate the hard-core dimer constraint.
Figure adapted from Ref.~\cite{albuquerque:08}.
}
\label{fig:jjk}
\end{figure}

The Josephson-junction array (JJK) can be described by the generalized
Bose-Hubbard model
\begin{equation}
{\mathcal H} = \frac{1}{2}\sum_{j,k}n_{j}\hat{C}^{-1}_{j,k}n_{k} - 
J_{h} \sum_{\left\langle j,k \right\rangle}(b^{\dagger}_{j}b_{k} 
+ b^{\dagger}_{k}b_{j}) ,
\label{eq:bose_hubb}
\end{equation}
where the positions of the X-shaped islands in the array are denoted
by the indices $j$ and $k$ and ${\left\langle j,k \right\rangle}$
represents nearest-neighbor pairs on the kagome lattice (see
Fig.~\ref{fig:jjk}). $n_{j} = b^{\dagger}_{j}b_{j}$ is the bosonic
occupation number at site $\vec{r_j}$, $J_{\rm h}$ is the Josephson
current between two X-shaped islands, and $\hat{C}$ is the array's
capacitance matrix. We restrict the analysis to the case of
hard-core bosons \cite{albuquerque:08}.

\subsection{Two-dimer flips}
\label{sec:flips}

\begin{figure}[!tbp]
\includegraphics[width=0.42\textwidth]{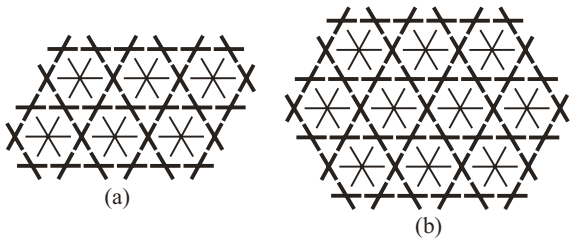}

\vspace*{-0.2cm}

\caption{
Open-boundary clusters studied: (a) $N\times2$ (here $N=3$)
hexagon ladders; (b) ten-hexagon cluster.
}
\label{fig:clusters}
\end{figure}

\begin{figure}[!tbp]
\includegraphics[width=0.9\columnwidth]{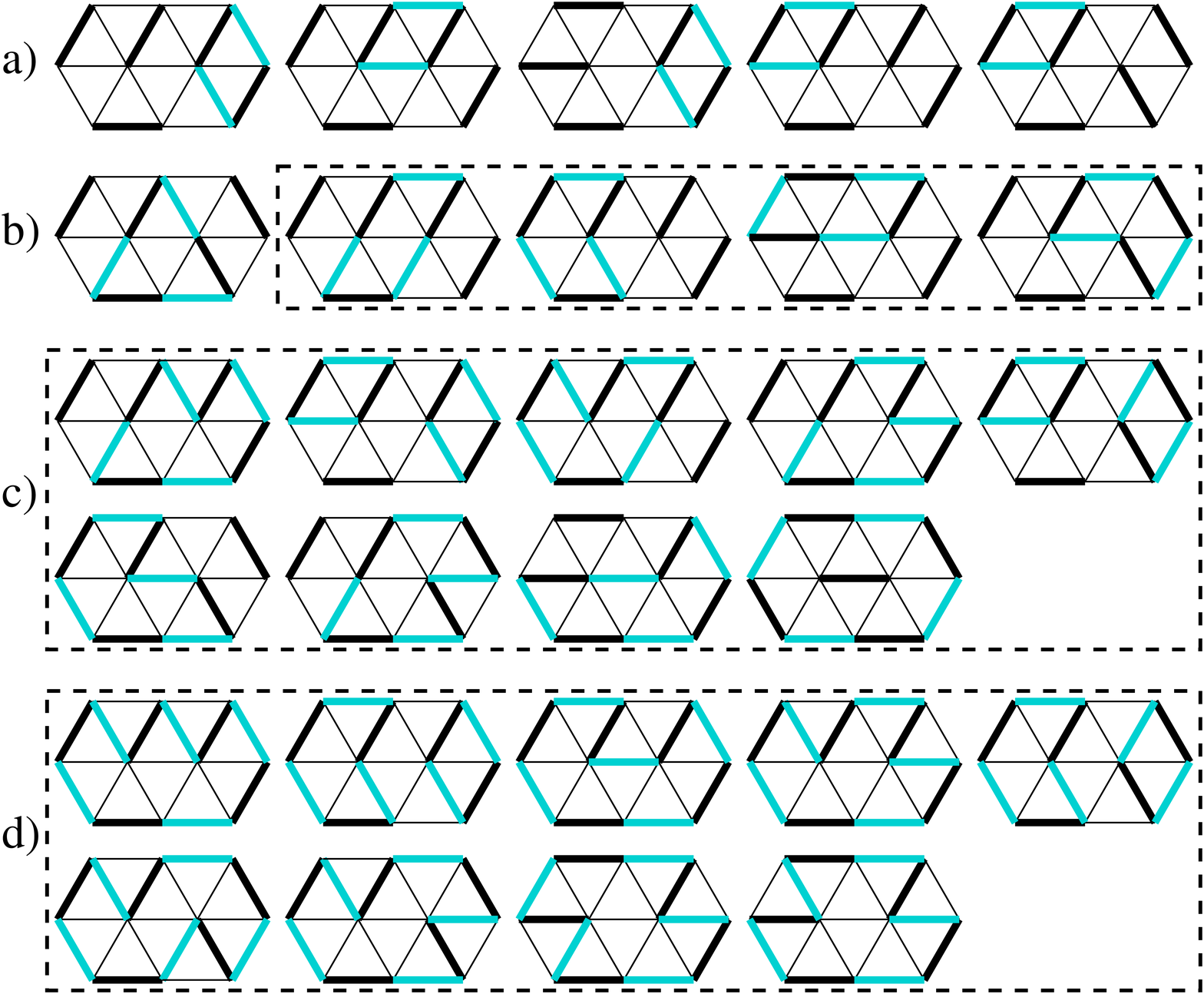}

\caption{(Color online)
Nonequivalent dimer flips in the ten-hexagon cluster
[Fig.~\ref{fig:clusters}(b)], comprising (a) two, (b) three, (c)
four, and (d) five dimers. Dimer flips are represented by their
associated {\em transition graph}: dimers (thick black lines) flip
to new positions (thick light lines) while observing the hard-core
constraint. Only the underlying triangular lattice of the JJK array
is shown (shaded lines in Fig.~\ref{fig:jjk}). The quantity $\Sigma$
used for gauging the validity of the mapping onto a QDM
is defined via the multi dimer flips enclosed by dashed lines.
}
\label{fig:flips}
\end{figure}

In this example we focus on the off-diagonal dimer flip term
$t$ in Eq.~(\ref{eq:qdm_plaq}) from the microscopic model
[Eq.~(\ref{eq:bose_hubb})] with the following set of capacitances:
$C_{\ast} = 1$, $C_{\rm X} = 0.25$, $C_{\rm i} = 2.5$, and $C_{\rm h}
= 0.5$ (see Fig.~\ref{fig:jjk}). Using ENCORE these are obtained
from Eq.~(\ref{eq:eff2}): the transition amplitude between dimer
configurations $\ket{\phi_m}$ and $\ket{\phi_{m'}}$ ($m \neq m'$)
is the matrix element ${\mathcal H}^{\rm eff} (m, m')$.  Our results
have been obtained on the clusters shown in Fig.~\ref{fig:clusters}.

The dominant off-diagonal term in the effective Hamiltonian is the
two-dimer flip with amplitude $t$. This process involves the creation
of a virtual state with a doubly-occupied hexagon, with energy
$E_{\rm hex}$, in the kagome lattice and occurs with amplitude $t
\approx J_{\rm h}^{2} / E_{\rm hex}$ \cite{ioffe:02,albuquerque:08}.
Two-dimer flips in the cluster with ten hexagons are shown in
Fig.~\ref{fig:flips}(a). Although the amplitudes for all these are
ideally equal, there are small deviations, e.g., by the configuration
of the neighboring dimers (effects of Coulomb interactions) or
the open boundaries. All two-dimer flips depicted in
Fig.~\ref{fig:flips}(a) can be seen as being {\em correlated}
and are considered individually at the algorithmic level.

\begin{figure}[!tbp]
\hspace*{-0.6cm}\includegraphics*[width=0.30\textwidth,angle=270]{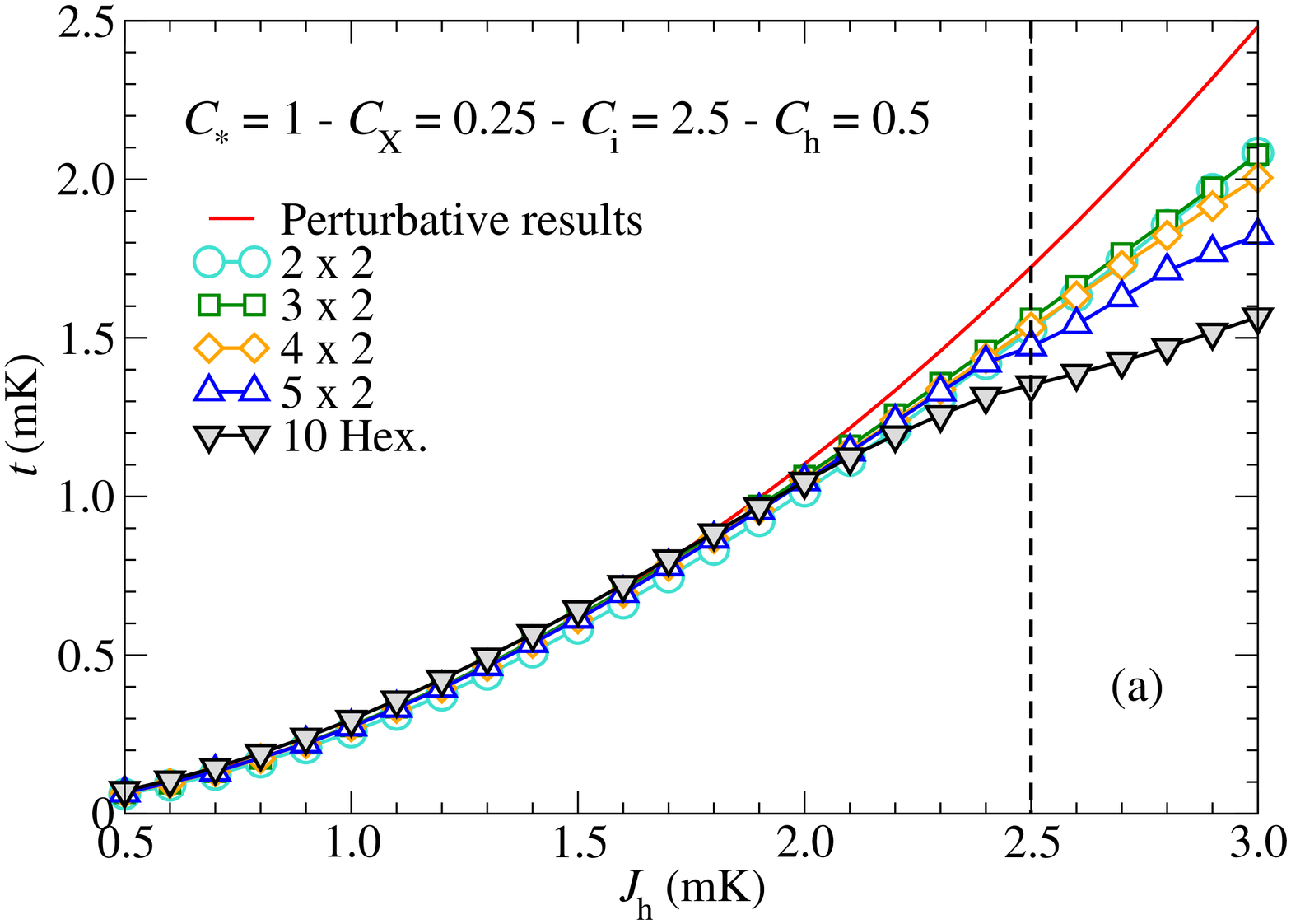}

\vspace{0.1cm}

\includegraphics*[width=0.3\textwidth,angle=270]{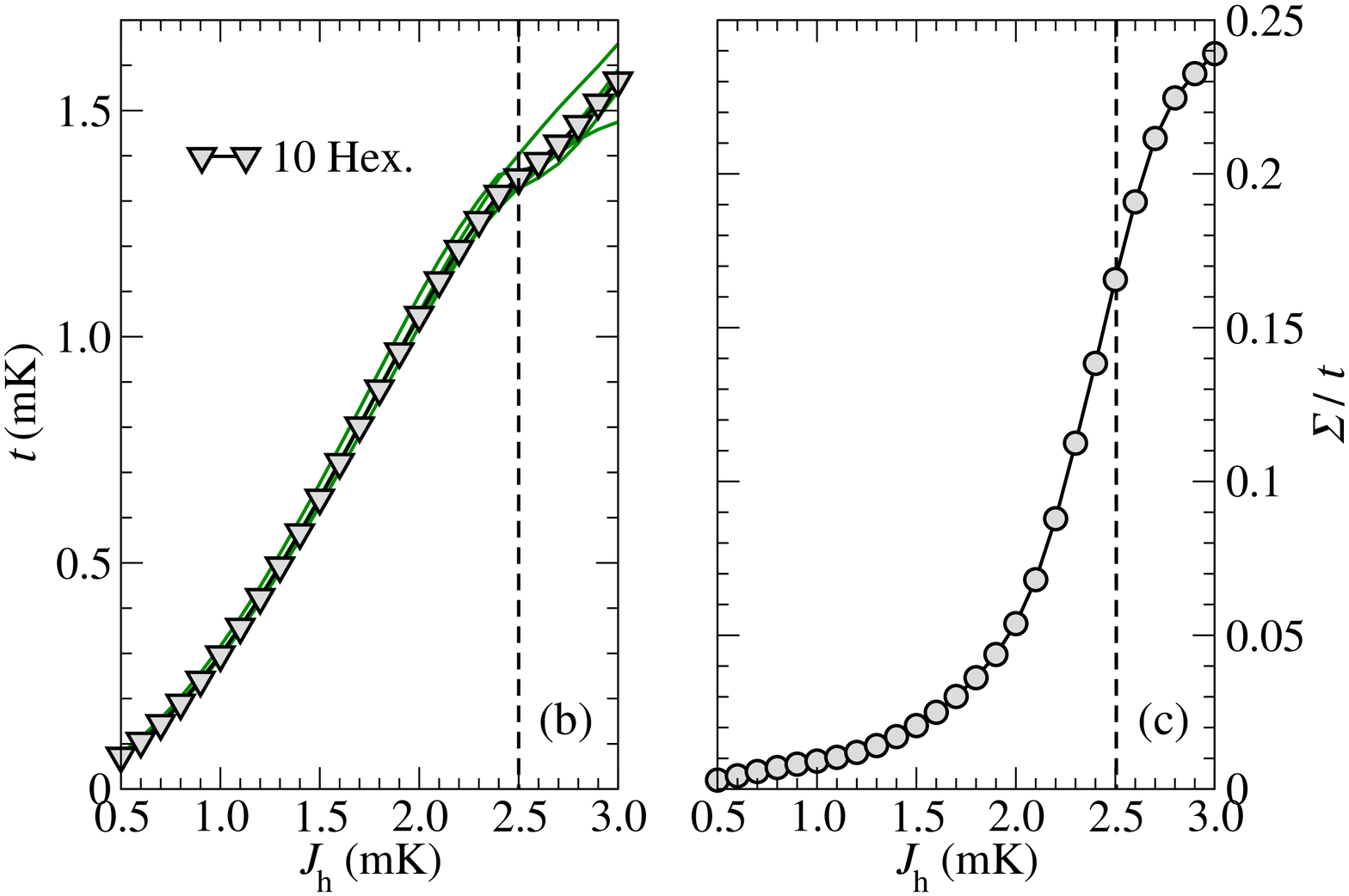}

\caption{(Color online) 
(a) Amplitude for the two-dimer flip $t$ in the JJK array
obtained from the ENCORE analysis of the finite clusters shown in
Fig.~\ref{fig:clusters}. The (red) solid curve represents second-order
perturbation results.  (b) Results for the ten hexagon cluster are
obtained as the average (triangles) of the amplitudes of the two-dimer
processes depicted in Fig.~\ref{fig:flips}(a).  (c) Added absolute values
for the amplitudes associated to multi dimer flips ($\Sigma$). When
$\Sigma$ is large the mapping onto the QDM breaks down (vertical dashed
lines). Data for $C_{\ast} = 1$, $C_{\rm X} = 0.25$,  $C_{\rm i} =
2.5$, and $C_{\rm h} = 0.5$ (adapted from Ref.~\cite{albuquerque:08}).
}
\label{fig:2_dimer_flip}
\end{figure}

Figure \ref{fig:2_dimer_flip}(a) shows results for $t$ as a function
of the Josephson coupling $J_{\rm h}$ for the various clusters and
in comparison to second-order perturbative results. Results for the
ten-hexagon cluster are obtained by averaging the amplitudes for the
processes depicted in Fig.~\ref{fig:flips}(a); amplitudes for the
individual processes are shown in Fig.~\ref{fig:2_dimer_flip}(b).
The results agree up to a point [vertical dashed lines in
Figs.~\ref{fig:2_dimer_flip}(a) -- \ref{fig:2_dimer_flip}(c)] where
the mapping onto a QDM fails. This agreement is an indication that,
for capacitances and Josephson currents leading to a valid mapping,
the low-energy physics of the JJK array is indeed described by a
QDM with local dimer resonances. Furthermore, it also points to the
absence of sizable finite-size effects in our results.

\subsection{Multi dimer flips: Breakdown of the mapping} 
\label{sec:multi_flips}

Whereas a standard CORE expansion proceeds by considering clusters
comprising an increasing number of {\em sites}, an ENCORE expansion for
the JJK array, due to the dimers' hard-core constraint, is performed
in terms of the number of {\em dimers} in a cluster.  The analysis
of multi dimer terms can be used to gauge the validity of the mapping
onto a QDM: large amplitudes for multi dimer flips indicate that the
device is not properly described by the effective model. We denote the
summed absolute value of the amplitudes associated to these multi dimer
flips by $\Sigma$, which are directly obtained as the off-diagonal
matrix elements of the effective Hamiltonian associated to the multi
dimer flips enclosed by dashed lines in Figs.~\ref{fig:flips}(b)
-- \ref{fig:flips}(d).  Figure \ref{fig:2_dimer_flip}(c) shows
$\Sigma$ as a function of the Josephson current $J_{\rm h}$.
A sudden increase in $\Sigma$ at the same value of $J_{\rm h}$ for
which different results for $t$ start to deviate from each other in
Figs.~\ref{fig:2_dimer_flip}(a) and \ref{fig:2_dimer_flip}(b) indicates
the breakdown of the mapping.  The appearance of ``intruder states''
in the low-energy spectrum with negligible overlap with the hardcore
dimer configurations also indicate the breakdown of the mapping.
The vertical dashed line in Fig.~\ref{fig:2_dimer_flip}(c) indicates
the point where the first intruder state appears.  As $J_{\rm h}$
increases and charge fluctuations start to dominate, intruder states
displaying multiply-occupied hexagons in the JJK array violating the
hard-core dimer constraint have their energy lowered, eventually
causing some of the projected states $\{ \ket{\psi_m} \}_{1}^{M}$
to vanish.

\section{Summary}

We have presented an ENCORE algorithm suitable for constrained
effective models whose basis states are not simply tensor products of
local block states. We find that CORE is very effective in the design
of quantum devices for emulating exotic phases. The inadequacy of
the restricted set of degrees of freedom in accounting for a system's
low-energy behavior is reflected by the presence of long-range terms
in the effective Hamiltonian obtained from CORE and is used as a
criterion in deciding on whether successful emulation is achieved.

\begin{acknowledgments}

We thank A.~Abendschein for fruitful discussions.  A.F.A.~acknowledges
financial support from CNPq (Brazil), NIDECO (Switzerland), and ARC
(Australia). H.G.K.~acknowledges support from the Swiss National
Science Foundation under Grant No.~PP002-114713. We would like to thank
the ETH Zurich Integrated Systems Laboratory and especially Aniello
Esposito for computer time on the large-memory workstation ``schreck.''

\end{acknowledgments}

\bibliography{refs,comments}

\begin{thebibliography}{21}
\bibitem[{com()}]{comment:solver}
\bibinfo{note}{Although the Lanczos \cite{lanczos:50} algorithm could in
  principle be used, it has problems with degenerate eigenvalues. Thus, for the
  smallest clusters a dense matrix eigensolver is recommended, whereas for
  larger clusters we use the Davidson algorithm \cite{davidson:75}.}
\end{thebibliography}

\end{document}